\documentclass{emulateapj}

\shorttitle{A pulsar wind nebula around \axp}
\shortauthors{J. Vink \& A. Bamba}

\begin{document}

\title{
The discovery of a pulsar wind nebula around the magnetar candidate
AXP 1E1547.0-5408}

\author{Jacco Vink$^1$ and Aya Bamba$^{2}$}
\affil{$^1$Astronomical Institute, University Utrecht, P.O. Box 80000, 
3508TA Utrecht, The Netherlands}

\affil{$^2$Dublin Institute for Advanced Studies,
5 Merrion Square, Dublin 2,
Ireland\\
$^3$ISAS/JAXA Department of High Energy Astrophysics,
3-1-1 Yoshinodai, Sagamihara, Kanagawa 229- 8510, Japan%\\
%DIAS, Dublin, Ireland
}
\email{j.vink@astro.uu.nl}

\def\cxo{{\it Chandra}}
\def\chandra{{\it Chandra}}
\def\xmm{{\it XMM-Newton}}
\def\adspr{{Adv. in Space Res.}}
\def\gcn{{GCN No.}}
\def\axp{1E\,1547.0-5408}
\def\kms{{km\,$s^{-1}$}}

\begin{abstract}
We report the detection of extended emission around 
the anomalous X-ray pulsar AXP 1E1547.0-5408 using
archival data of the {\em Chandra} X-ray satellite. 
The extended emission consists of an inner part, with an extent
of 45\arcsec\ and an outer part with an outer radius of 2.9\arcmin, which coincides
with a supernova remnant shell  previously detected in the radio.
We argue that the extended emission in the inner part is the result
of a pulsar wind nebula, which would be the first detected pulsar wind
nebula around a magnetar candidate.
Its ratio of X-ray luminosity versus pulsar
spin-down power is comparable to that of other young pulsar wind nebulae,
but its X-ray spectrum is steeper than most pulsar wind nebulae.
We discuss the importance of this source in the context of magnetar evolution.
\end{abstract}


\section{Introduction}

Among the various types of neutron stars, soft gamma-ray repeaters (SGRs) and
anomalous X-ray pulsars (AXP) stand out by their extreme bursting behavior,
high X-ray luminosity, and long rotation periods of 2-12 s
\citep[see][for reviews]{woods04,mereghetti08}. 
They are now widely recognized as magnetars, 
neutron stars with extreme surface magnetic fields
of $10^{14}-10^{15}$~G, well above the quantum critical magnetic field
of $B_{QED}= 4.4\times 10^{13}$~G.
The X-ray luminosity and bursts of magnetars are thought to be powered by
the decaying magnetic field. Despite their extreme energy output, the
particle acceleration properties of magnetars are not well known.

Here we report the detection of a 
pulsar wind nebula (PWN) around AXP \axp.
We show that the nebula's brightness is
comparable to that of normal pulsars with the same rotational 
energy loss rate. This suggests that the nebula is powered by 
rotational energy loss, rather than by the magnetar's magnetic field.

\begin{figure*}
%\plottwo{f1a.ps}{f1b.ps}
\includegraphics[width=0.5\textwidth]{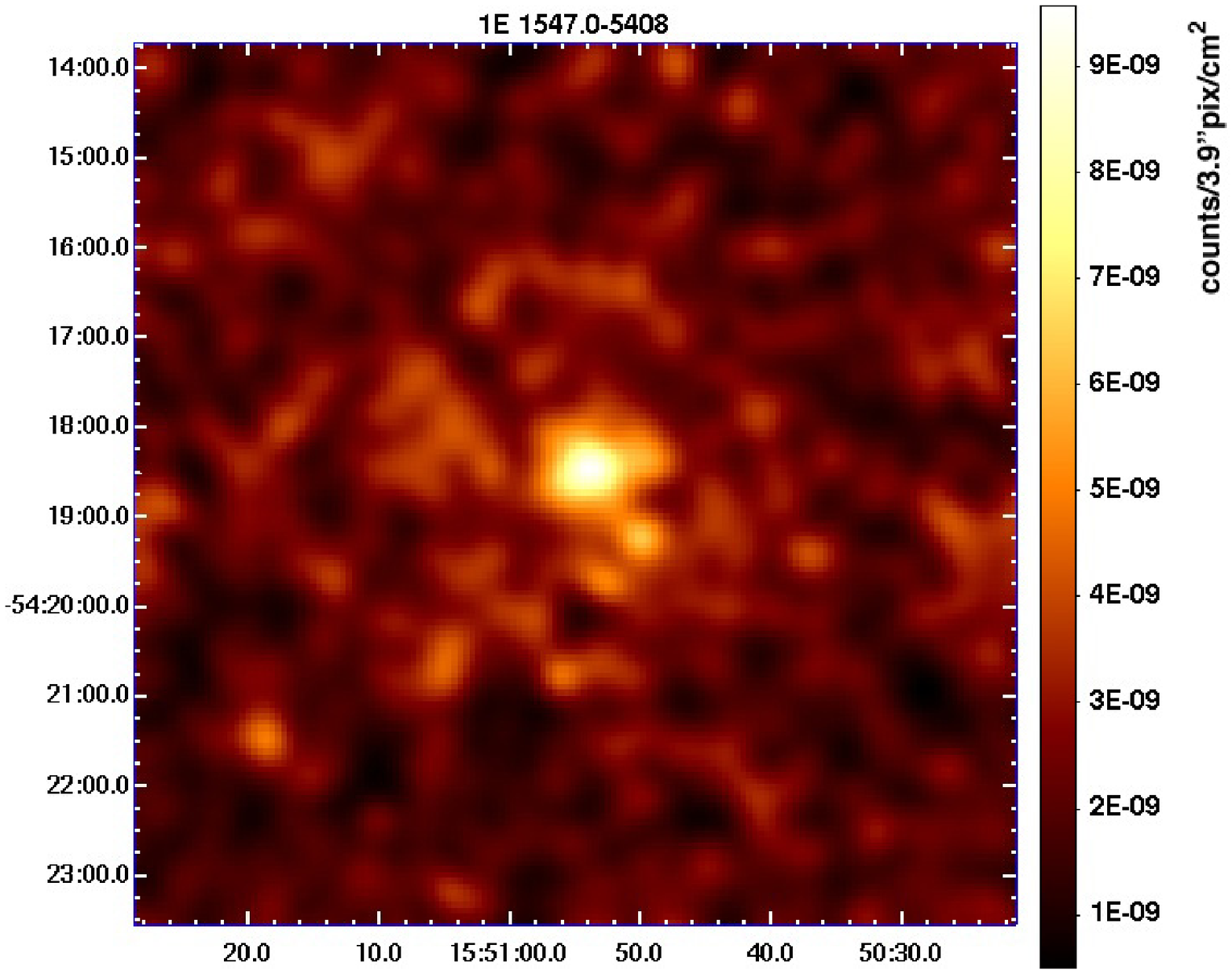}
\includegraphics[width=0.5\textwidth]{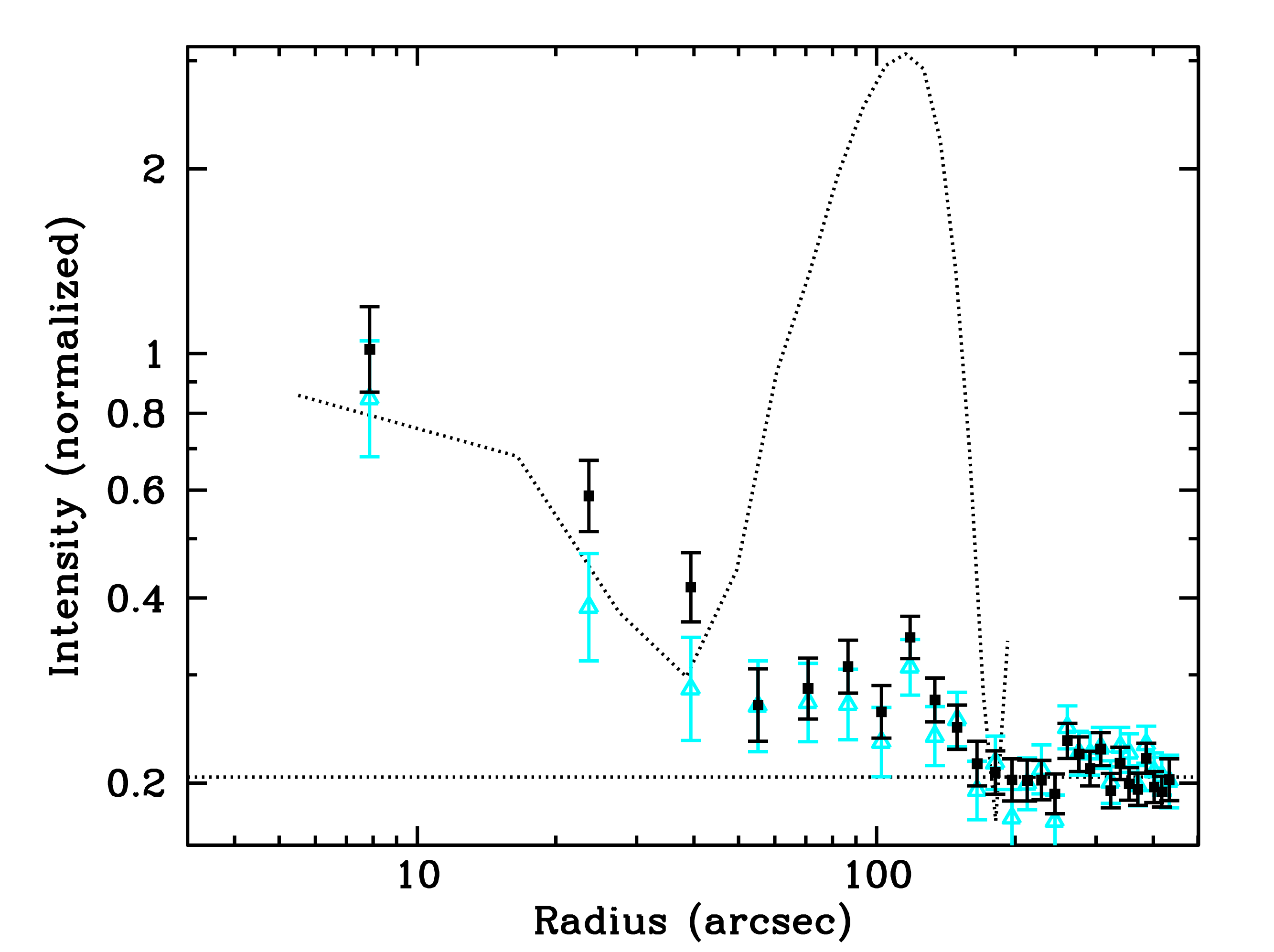}
\caption{
Left: 
A \chandra\ ACIS-I image of the diffuse emission
surrounding \axp\ in the energy range 1.5-7 keV.
Several point sources, including \axp, were removed, and the image
was then smoothed with a kernel of $\sigma = 11.8$\arcsec\ (24 pixels).
Right: The radial brightness profile in the 1.5-7 keV band (black squares) 
and radio
(dotted line, based on the SUMMS survey \citep{bock99}).
We have normalized the radio profile such that the central enhancement coincides
with extended emission in X-rays.
The X-ray background intensity is indicated by the dashed line, and
the spectral extraction regions for the PWN, shell and background (bgd) 
by two-sided arrows.
With cyan triangles we also indicate the profile in the 3-7 keV range.
\label{fig_imaging}
}
\end{figure*}

\subsection{\axp}
\axp\ \citep{lamb81}
is an X-ray source that was only recently recognized to
belong to the class of magnetars \citep{gelfand07}, 
a name we adopt here to refer to both SGRs and
AXPs. With its rotation period of 
$P=2.1$~s 
\citep{camilo07}, it is the most rapidly spinning magnetar known.
Its period derivative is $\dot{P}=2.3\times 10^{-11}$, which combined
with the period, implies a  surface magnetic field of 
$B_{dipole}=3.2\times 10^{19}\sqrt{P\dot{P}}=2.2\times 10^{14}$~G, and a rotational energy loss rate of 
$\dot{E} = 1.0\times 10^{35}$~erg\,s$^{-1}$, the highest among 
AXPs and SGRs.
\axp\ is also a transient radio source
\citep{camilo07},
a characteristic shared with only one other magnetar, 
AXP XTE J1810-197 \citep{camilo06}.
Very recently \axp\ went through a period of major bursting activity 
\citep{krimm08a,krimm08b,gronwall09}, confirming its
status as a magnetar.

\axp\ is located inside a supernova remnant (SNR)
shell, G327.24-0.13, detected with the Molongo radio observatory
%at 843 MHz and 1.4 GHz
 \citep{gelfand07}. 
Although the spin-down rates of most AXPs/SGRs imply that they are young
sources (typically $< 10^5$~yr), only a few of them are 
associated with SNRs \citep{gaensler04}.
These SNRs may hold important clues about the supernovae
that created the magnetars. In particular, the energy of the
remnants indicates that the supernovae creating magnetars are not
more powerful than ordinary supernovae \citep{vink06c}.
This poses a challenge to the scenario in which magnetars are born
with extremely high initial spin-periods \citep{duncan92}, 
in which case $\sim 10^{52}$~erg rotational energy is likely to 
be transfered to the supernova ejecta. 
Nevertheless, this scenario cannot be completely excluded, because
the distances used by \citet{vink06c} have been disputed
\citep{durant06}.

A major obstacle in determining the initial spin-periods of AXPs/SGRs
is that 
they spin-down rapidly and often
irregularly \citep[e.g.][]{gavriil04,esposito09}. It is, therefore,
difficult to determine whether they were born with periods close
to their current spin-periods, or 
with periods comparable to, or even shorter than ordinary pulsars.
Provided that the nebula around \axp\ is indeed powered by
rotational energy loss, its properties may help to clarify its rotational
history, as PWNe contain most of the initial rotational energy lost over
the lifetime of the pulsar \citep[e.g.][]{pacini73}.

In the next section 
we report on the details of the detection of extended emission around \axp,
both based on imaging (\S~\ref{sec_image}), 
and on spectral analysis (\S~\ref{sec_spec}). In \S~\ref{sec_dust}
we argue%, based on the spectral data, 
that the extended source is most likely a PWN, 
and not a dust scattering halo. In the final section we
discuss the implications of the presence of a PWN around \axp.

\section{Data analysis and results}
\subsection{Data selection}
In order to find a potential X-ray counterpart to the radio shell
of G327.24-0.13, we analyzed archival \chandra\ and \xmm\ data.
During this search we found a faint, but distinct, extended X-ray source, 
centered on the magnetar. Not all archival data were suitable
for our analysis, since several \chandra\ observations were taken in
continuous clocking mode, which does not allow for imaging.
This leaves only the 9.6~ks long 
observation on July 1, 2006, with the ACIS-I detector,
as a suitable observation for extended source analysis.
Several \xmm\  observations are present in the archive, but
the point spread function (PSF)
of \xmm\ is much broader ($\sim 6$\arcsec\ FWHM, with extended broad wings) 
than that of \chandra\ ($\sim 0.5$\arcsec\ FWHM).
Therefore, the \chandra\ data, although poor in statistics, provide
a more convincing case for the presence of an extended source. 
However, we note 
that the \xmm\ data are 
consistent with the results presented here.

\subsection{Imaging}
\label{sec_image}
We performed the analysis of the \chandra\ data using the standard
software (CIAO v4.1) to make exposure corrected images, and
extract spectra of the magnetar,
the extended emission, and the region of the radio shell that was
found by \citet{gelfand07}.
The extended source is %very 
weak, and can only be seen after
significantly smoothing the images. However, the advantage of the
small PSF of \chandra\ is that point sources can be effectively
removed before smoothing. 
Apart from X-ray emission within 4\arcsec\ of the position
of  \axp\ also other faint point sources in the field were removed,
two of which were located in the region of the radio shell\footnote{
Their J2000 positions are $\alpha=15^h51^m07.7^s, \delta=-54\degr 19\arcmin 
26.0\arcsec$, 
and $\alpha=15^h50^m34.8^s, \delta=-54\degr 19\arcmin 25.9\arcsec$.}.
The resulting 1.5-7 keV image 
is shown in Fig.~\ref{fig_imaging} (left). We chose for this energy range, 
because below 1.5 keV there is hardly any emission to be expected due to to
de large absorption column \citep[c.f. Fig.~2 in][]{gelfand07}.
The image is noisy, but extended emission centered on
\axp\ can be seen. This source is surrounded by a faint shell.

The radial X-ray brightness profile makes the structure even more
clear (Fig.~\ref{fig_imaging} right).
We include in this figure the radio intensity profile at 843 MHz as obtained
from the Sydney University Molonglo Sky Survey \citep{bock99,gelfand07}.
We also show the profile using the hardest photons only,
3-7 keV, as this lowers the potential contribution of
dust scattered X-ray emission from \axp\ 
(see \S~\ref{sec_dust}).
The profile clearly shows the presence of an extended emission
component that
falls off with radius and disappears at 
$r \approx 45$\arcsec. Further outward there is 
a small enhancement out to
$r \approx 150$\arcsec, which coincides
with the radio shell.
Also the radio map gives an hint
of centrally enhanced emission  with 
an approximate flux
density of  $\sim$0.008 Jy at 843 MHz and  a radial profile similar
to that in  X-rays, but with a different brightness ratio between
shell and central emission. 

The radial brightness profiles allow us to make an estimate of the statistical
detection significances. Fitting the first three
bins, labeled PWN, with a flat background model results
in $\chi^2/{\rm d.o.f}=55.0/3$, corresponding
to a 6.9$\sigma$\ detection. In the 3-7 keV band this reduces to 
$\chi^2/{\rm d.o.f}=15.3/3$, still a $3.2\sigma$ detection. For the
radio shell the numbers are $\chi^2/{\rm d.o.f}=58.6/7$ ($6\sigma$)
and $\chi^2/{\rm d.o.f}=18.1/7$  ($2.5\sigma$), respectively.

\begin{figure}
\includegraphics[width=\columnwidth]{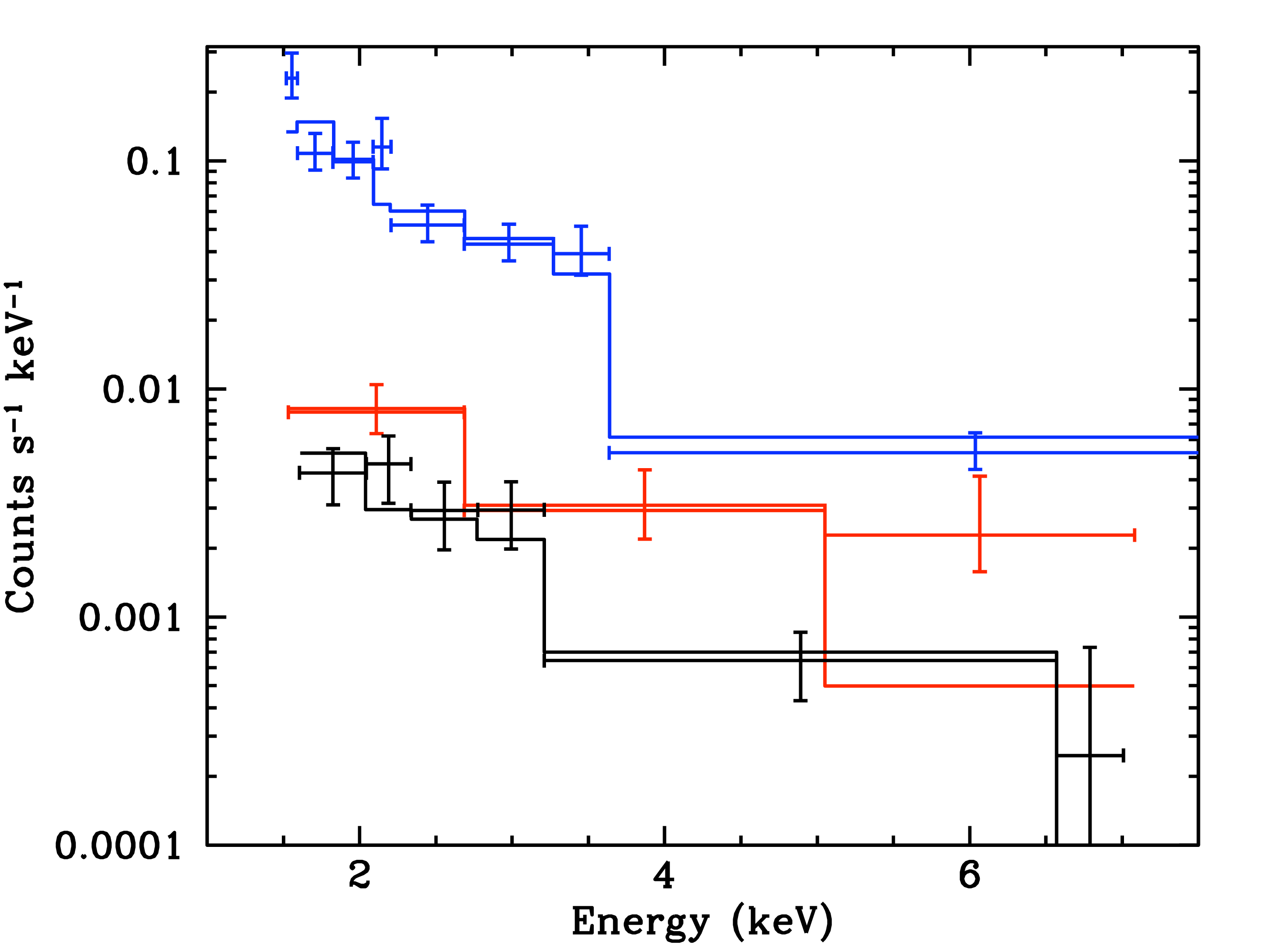}
\caption{
The \chandra\ X-ray spectra of the pulsar wind nebula, 
the supernova remnant shell and \axp.
The countrates of \axp\ and the pulsar wind nebula have been multiplied
by the factors indicated for reasons of clarity. The rebinning was
done for display purposes. No rebinning was necessary for the fitting
itself as the statistic described in \citet{cash79} was used.
\label{fig_spectra}
}
\end{figure}

\begin{table}
\begin{center}
\caption{Spectral fit parameters\tablenotemark{a}
\label{table_spec}} 
\begin{tabular}{llll}\tableline\tableline\noalign{\smallskip}
Component & \multicolumn{2}{c}{Flux 
 (2-10 keV)}      & Spectral index  \\
          & \multicolumn{2}{c}{($10^{-13}$ erg s$^{-1}$cm$^{-2}$)}   \\
          & absorbed\tablenotemark{b} & unabsorbed & \\\noalign{\smallskip}
\tableline\noalign{\smallskip}
AXP                & $2.4 \pm 0.3$ & $3.8 \pm 0.4$ & $-3.8 \pm 0.1$\\
Extended source & $1.0 \pm 0.3$ & $1.5\pm 0.3$  &    $-3.4 \pm 0.4$ \\
Shell              & $2.3 \pm 0.7$ & $3.5\pm0.7$ &   $-3.3 \pm 0.8$ \\
\noalign{\smallskip}
\tableline
\end{tabular}
\tablenotetext{a}{Based on joint fits of  the spectra in
the 1.5-7 keV range using the {\em xpsec} v11 package, and
employing the C-statistic, \citep{cash79}. 
The overall statistic for the joint fit was $C=74.5$ for 72 bins.
Excluding the PWN model fit resulted in $\Delta C=125$ ($11\sigma$), and
excluding the Shell model resulted in $\Delta C=110$ ($6\sigma$).
}
\tablenotetext{b}{The best fit absorption column was
$N_{H} = (2.75\pm0.25)\times 10^{22}$~cm$^{-2}$, using the {\em tbabs} absorption model \citep{wilms00}.
}
\end{center}
\end{table}

\subsection{Spectral analysis}
\label{sec_spec}
We extracted X-ray spectra of three regions, using a radius of $r=0.74$\arcsec\
to isolate the spectrum of \axp\ (corresponding to 80\% encircled
energy);
an annulus with inner radius of 4\arcsec\ and an outer radius of
$45$\arcsec\ for the extended source;
and an annulus between 45\arcsec\
and  $174$\arcsec\ for the X-ray counterpart of the
radio shell. The inner radius for the extended source
guarantees that only $\sim 5$\% of the point source contributes 
to the emission of the
extended source, due to the scattering wings of the \chandra\ PSF.
\footnote{The \chandra\ Proposers' Observatory Guide V11.0 Fig.~4.20 shows that
95\% of the photons with energies of 1.5 keV fall within 2\arcsec\ of 
the source position.
}
Given that the flux of \axp\ appears to be about twice as large as
of the extended source (Table~\ref{table_spec}), 
we expect a $\sim 10$\% contamination of the extended source from the AXP,
due to the wings of the PSF.
The background spectrum of the extended source and the shell was obtained
from an annulus centered on the AXP with $180$\arcsec\ $< r < 272$\arcsec. 
The background spectrum for the AXP was obtained from
an annulus with 4\arcsec\ $<r<20$\arcsec,%\ surrounding the point source,
i.e. we treated the extended region as a source of background for the point 
source.

We fitted these data with power law spectra with a Galactic absorption
component. 
For our analysis we also investigated an additional 
black-body component for the point source, and a non-equilibrium ionization
thin plasma model for the radio shell. However, 
this did not improve the fits. Note that for the \xmm\ data including a 
black-body for the point source did improve the fit \citep{gelfand07},
but the \xmm\ data has better statistics concerning the spectrum of  the point 
source. According to \citet{gelfand07} the best spectrum of the AXP is
a combination of a soft power law with index -3.7 plus a black-body component
with $kT=0.43$~keV. The \chandra\ data agrees with this in the sense
that we also find a steep power law.
The resulting spectra are shown in Fig.~\ref{fig_spectra} and the best
fit parameters
are listed in the Table~\ref{table_spec}.

\begin{table}
\begin{center}
\caption{Expected dust scattering contributions\tablenotemark{a} 
\label{table_dust}} 
\begin{tabular}{lllll}\tableline\tableline\noalign{\smallskip}
$\tau_{sca}$ & $\beta$\tablenotemark{b}
& $F_{halo}/F_{obs}$
&  $F_{halo}/F_{obs}$
& $\Delta \Gamma$ \\\smallskip
   (@ 1 keV)  &       &1.5~keV        & 3 keV             & \\\noalign{\smallskip}
\tableline\noalign{\smallskip}
1.5 & 0 & 13\% & 6\%  & -0.4\\
1.5 & 1 & 25\% & 11\% & -0.5\\
\noalign{\smallskip}
3.0 & 0 & 37\% & 13\%  & -0.9\\
3.0 & 1 & 73\% & 25\%  & -0.9\\
%\noalign{\smallskip}
%3.5 & 0 & 51\% & 16\%  & -1.0\\
%3.5 & 1 & 100\% & 30\%  & -1.1\\
\noalign{\smallskip}
\tableline
\end{tabular}
\tablenotetext{a}{Prediction based on Eq.~19
of \citet{draine03}.}
\tablenotetext{b}{
$\beta=0$ corresponds to uniformly distributed gas,
$\beta=1$ has 3/4th of the dust at a distance greater than 50\% of the
total distance \citep{draine03}.
}

\end{center}
\end{table}

\subsection{A pulsar wind nebula or a dust scattering halo?}
\label{sec_dust}
The spatial morphology and the spectral characteristics of the extended source
have all the characteristics of a PWN created by \axp,
but we need to exclude the possibility that 
the extended emission could be caused
by a dust scattering halo \citep[e.g.][]{predehl95}. 
Indeed, dust scattering is not absent as indicated by 
the recent detections of 
evolving dust scattering haloes after
some major outbursts of \axp\ \citep{tiengo09}. 
However, in the situation investigated here
a dust scattering halo is unlikely to dominate the emission above 3~keV,
with a most likely flux contribution to the emission within 45\arcsec
of 6-11\%.

Our estimate is based on the relation between the dust scattering 
optical depth $\tau_{sca}$ at 1~keV and the absorption column
as found by \citet{predehl95}. For 
$N_{\rm H} = 3\times 10^{22}$~cm$^{-2}$ 
\citep[Table~\ref{table_spec} and][]{gelfand07} this gives 
$\tau_{sca}=1.5$ at 1 keV, corresponding to $\tau_{sca}=0.17$ at
3 keV, because the dust scattering cross section is proportional to
$E^{-2}$. The values of $\tau_{sca}$ can be related to the fractional
halo intensity,  $I_{frac}$ \citep{predehl95}. 
At 3~keV this gives $I_{frac}=15$\%, 
but most of the halo flux will be outside the spectral extraction radius 
of 45\arcsec. To estimate the
fraction within 45\arcsec, we use the semi-analytic approximation
of \citet{draine03}, which employs a parameter $\beta$ to account for
gradients in the dust distribution; with $\beta=-1 (1)$ corresponding to a dust
distribution skewed in distance toward the observer (source), 
$\beta=0$ to a uniform
distribution. Table~\ref{table_dust} lists
the potential dust halo contributions to the extended source, 
which indicates a modest contribution of 11\% even for $\beta=1$.

\citet{draine03} discussed the possibility that the relation
$N_{\rm H}$ versus $\tau_{sca}$ relation of \citet{predehl95}
underestimates $\tau_{sca}$, 
due to the neglect of small angle scattering. For that reason we
also provide in Table~\ref{table_dust} values for the expected dust halo
contributions for $\tau_{sca}=3.0$. In that case the dust halo becomes more
substantial, but at 3~keV, with 25\% for $\beta=1$, it still not dominant.
Note that the model by \citet{draine03} seems to
overestimate dust halo scattering: 
Apart from its discrepancy with the results of \citet{predehl95}, also
a recent \chandra\ observation 
of GX13+1 shows that this model gives a poor fit to the data and
underestimates the absorption column \citep{smith08}, and overestimates
the fraction of small angle scattering. 
This suggest that the 25\% contribution
at 3 keV is too conservative, and an $<11$\% seems likely.
Moreover, adopting
$\tau_{sca}=3.0$ predicts a steepening of the halo spectrum within
45\arcsec\ with respect to the source spectrum of $\Delta \Gamma=-0.9$,
whereas the actual spectrum of the extended source is flatter than that
of the AXP by $\Delta \Gamma=0.4\pm 0.5$\ (a 2.7$\sigma$ difference). 
Note that the expected steepening,
is determined by the interplay between the energy
dependence of the cross section, $\sigma \propto E^{-2}$, and the energy
dependence of the scattering angles. 
%(at lower energies the halo is more extended). 
Still higher values
of $\tau_{sca}$ result in even stronger spectral steepening.

\begin{figure}
\includegraphics[angle=-90,width=\columnwidth]{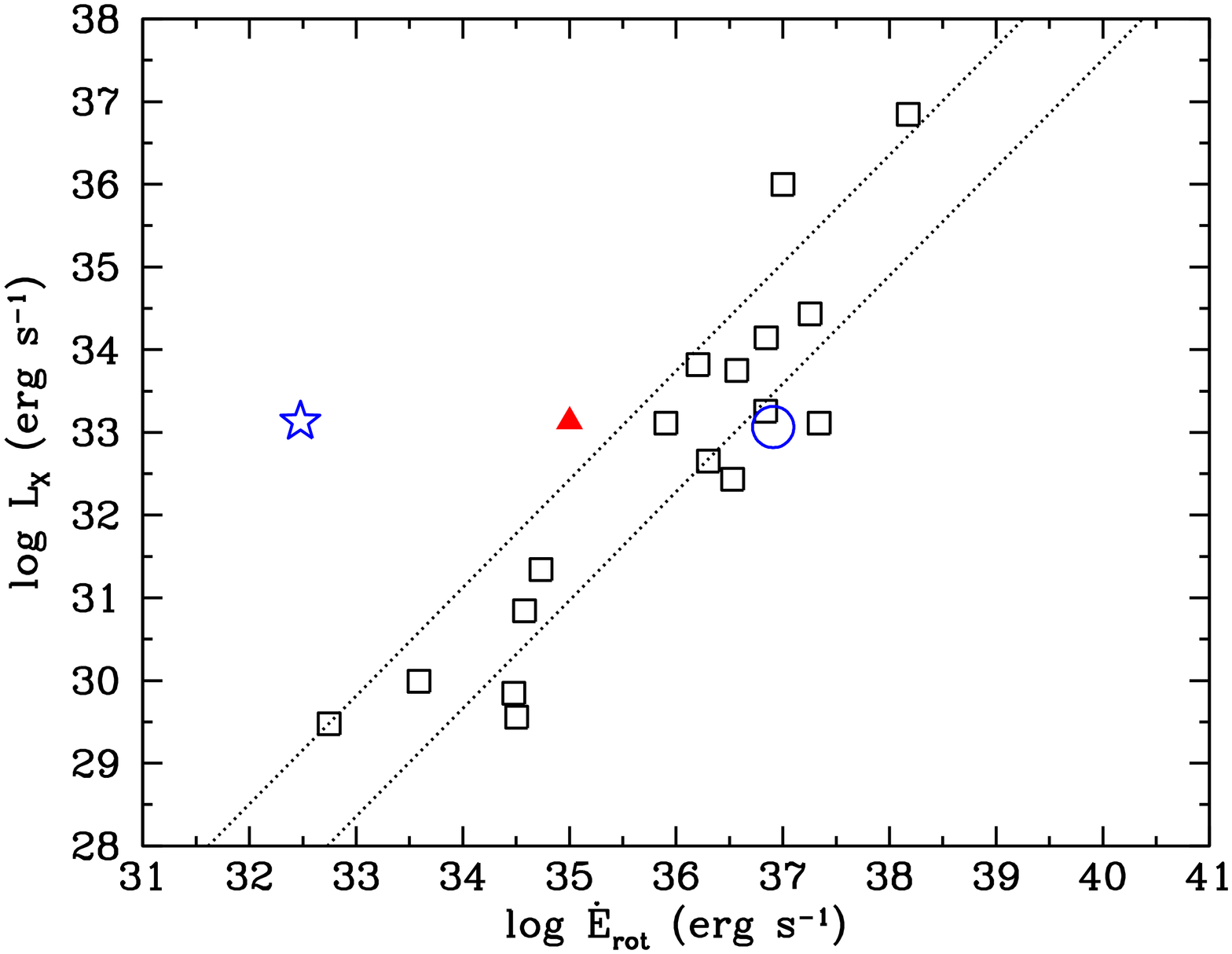}
\caption{
The X-ray luminosity, $L_X (2-10 keV)$, and the rotational energy loss,
$\dot{E}_{rot}$ of the magnetar \axp\ (filled triangle)
compared to the unpulsed X-ray emission of other X-ray pulsars \citep{cheng04}.
A distance of 9~kpc for \axp\ is assumed \citep{camilo07}.
The dashed lines indicate the standard deviation around the linear
regression relation between $\dot{E}_{rot}$ and $L_X$.
Also indicated are the high magnetic field pulsars 
PSR J1846-0258 \citep{kargaltsev08,gavriil08} and PSR J1119-6127
\citep{safi-harb08} with filled circles, and
RRAT J1819-1458 \citep{rea09} with a star-like symbol.
\label{fig_pulsars}
}
\end{figure}

\section{Discussion and conclusions}
We report here the presence of an extended source around \axp,
which is most likely a PWN.
Not counting the spurious \citep{kaplan02d} association of
a radio nebula with SGR 1806-20 \citep{kulkarni94}, this is the first
evidence for a PWN around  a magnetar.
This is of particular interest, 
as young PWNe are formed during the whole lifetime of the pulsar,
and thus hold a record of the
energy released over the lifetime
of the pulsars \citep{pacini73,vanderswaluw01,chevalier04}.
This is important for magnetars,
because according
to one theory magnetars are born with rapid initial
spin periods ($P_i \lesssim 3$~ms) \citep{duncan92,duncan96},
a theory that was recently challenged on observational grounds
\citep{ferrario06,vink06c} .
But even if initial periods shorter than 3~ms seem unlikely, little
is known about the initial spin-periods of magnetars.
The presence of a PWN may therefore cast new light on this issue,
as it may provide evidence 
that the initial spin period was much shorter than the present spin-period.

In addition, the PWN around \axp\ may provide new insight into 
the connection between
AXP/SGRs and other young
pulsars that have slightly lower magnetic fields
\citep{mclaughlin03,kaspi05}, 
and that do not have all the magnetar characteristics.
%, such as a
%bright X-ray luminosity and/or bursting behavior.
A recent AXP-like burst from the rotation powered pulsar
PSR J1846-0258 already suggests that there is a continuum
in properties from pulsars with average magnetic fields of $10^{12}$~G
to magnetars \citep{gavriil08}.  PSR J1846-0258 has an inferred 
surface magnetic field of $4.9\times 10^{13}$~G and  
is surrounded by a PWN
and a SNR shell, Kes 75 \citep{kargaltsev08}.
Another pulsar with a relatively high surface magnetic field
 ($B=4\times 10^{13}$~G), PSR J1119-6127, is also surrounded by a PWN, 
which appears jet-like in morphology \citep{safi-harb08}. 
A third case of a high magnetic field pulsar with a PWN,
RRAT J1819-1448, was recently reported by \citet{rea09}.

One problem in interpreting the presence of a PWN around
a magnetar is that we do not a priori know whether the
PWN is powered by rotational energy loss, as is the case for normal pulsars,
or whether the nebula is  ultimately powered by the magnetic field 
of the magnetar.
For example it has been suggested that
magnetar magnetospheres are filled with electrons/positrons and/or ions 
\citep{thompson02}, a fraction of which may
escape in the form of a relativistic particle wind 
\citep{harding96,harding99}.

A hint that the PWN around \axp\ 
may be powered by the pulsar spin-down is provived
by comparing the ratio of the X-ray luminosity over the spin-down power 
of the PWN. 
There is a correlation between the X-ray luminosity
and spin-down power of normal young pulsar \citep{seward88,verbunt96,cheng04},
with $\eta_X = L_X/\dot{E}_{rot}$ in the
range of $10^{-4} - 10^{-2}$. 
Fig.~\ref{fig_pulsars} shows the correlation as taken from 
the sample of \citet{cheng04}. 
It shows that the PWN around \axp\ is consistent
with the general trend, with $\eta \approx 0.01$, 
although perhaps with a somewhat enhanced
X-ray luminosity compared to pulsars with a similar rotational energy loss
(at the 1.5$\sigma$ level). This is unlike the case of RRAT J1819-1448, which
has $\eta = 0.2$ \citep{rea09}.
The value of $\eta \approx 0.01$ for \axp\  
is consistent with the idea that its PWN is powered by
the pulsar's spin-down.
Interestingly,
this can be taken as an additional 
argument against the alternative theory for
the behavior of AXP/SGRs %, in particular their strong spin-down rate, 
namely that they are not magnetars, but neutron stars that have a high spin-down
rate due the propeller mechanism \citep[e.g.][]{marsden01}.

The idea that the PWN is powered by the pulsar spin-down is also credible
given the fact that 
\axp\ has the largest spin-down power among AXPs/SGRs.
If the PWNe around magnetars are powered by
magnetic activity, it is not clear why other AXPs/SGRs, some even more
active than \axp, do not show evidence for extended emission.
On the other hand, it is not quite clear why the X-ray spectrum of the PWN is 
rather steep, $\Gamma\approx -3.5$, compared to PWNe around ordinary pulsars,
which have $\Gamma\approx -2$ \citep{kargaltsev08}.
This may be taken as an hint for another origin, but it may also be explained 
by the much faster spin evolution of a magnetar compared to other young pulsars.
Using the terminology of \citet{pacini73}, most young PWN are in phase 2 of 
their evolution, i.e. the
pulsar has been a more or less constant source of relativistiv particles.
In that case the spectral index beyond the synchrotron break is  
steeper by only $\Delta \Gamma=0.5$
compared to the radio spectral index. 
The PWN around \axp\
is most likely in phase 3: its spin-down luminosity has been rapidly
 decreasing during its short life ($\tau_{char}\sim 1400$~yr).
This means that the average age of the relativistic electron populations is 
skewed toward the total age of the PWN. However, this does not necessarily lead
to a steeper spectrum \citep[e.g.][]{pacini73}. This is an
aspect of the nebula around \axp\ that  needs further clarification.

Finally, we emphasize that our discovery is based on a 
\chandra\ observation of only 9.6~ks. A much deeper observation is likely to
reveal further details. 
For the origin of the PWN, it will be important to know whether
there is any subtructure that may indicate that the nebula has been created as 
the result
of a discrete number of flares, whether there are jets,
or whether it is a more homogeneous source. Deeper observatons
are also important to obtain better spectra and images of the SNR.
This may reveal the nature of its X-ray emission, thermal, or non-thermal,
which will help to put better constraints on the age of the SNR.
In our view \axp\ and the extended sources around it may be key objects
to understand the origin and evolution of magnetars.

\acknowledgments
We thank Elisa Costantini for discussions on interstellar dust 
scattering, and Frank Verbunt for his careful reading of 
the manuscript. JV is supported by a NWO Vidi grant.

%\bibliographystyle{apj}
%\bibliography{ns,magnetar,dust,snrs,pwn}

\end{document}